\documentstyle[prl,aps,preprint]{revtex}

\newcommand{\be}{\begin{equation}}
\newcommand{\ee}{\end{equation}}
\newcommand{\half}{\frac{1}{2}}

\newcommand\undertilde[1]{\mathop{#1}\limits_{\widetilde{}}}
\newcommand{\fract}[2]{{\textstyle\frac{#1}{#2}}}
\begin{document}
\draft

\setcounter{page}{0}


\preprint{\vbox{MIT-CTP-3171 \null\hfill\rm January 1, 2002} }


\title{\vspace*{1pc}
\large\bfseries Delocalization of the axial charge in the chiral
limit}

\author{ R.L.~Jaffe\\[2ex]}

\address{Center for Theoretical Physics\\
  Laboratory for Nuclear Science
  and Department of Physics\\
  Massachusetts Institute of Technology\\
  Cambridge, Massachusetts 02139\\
  and\\
  RIKEN-BNL Research Center\\
  Brookhaven National Laboratory\\
Upton, New York 11973}

\maketitle
\thispagestyle{empty}
\begin{abstract}\noindent The nucleon's axial vector charge, $g_{A}$,
becomes
delocalized in the chiral limit.  When $m_{\pi}=0$, and
$SU(2)_{L}\times SU(2)_{R}$ is exact, $1/3$ of the nucleon's axial
charge is to be
found at infinite distance from the nucleon.  For finite $m_{\pi}$
this result is
approached smoothly as $m_{\pi}\to 0$. We illustrate this effect by
considering
the lepton-proton spin-spin interaction arising from $Z^{0}$ exchange
as a
function of $m_{\pi}$. Delocalization may have implications for
lattice
calculations of $g_{A}$ and in nuclei.
\end{abstract}


\narrowtext \newpage In this Letter I point out that the nucleon's
axial vector charge,
$g_{A}$, becomes delocalized in the chiral limit.  Specifically, when
$m_{\pi}=0$, exactly $1/3$ of the nucleon's axial charge is to be
found at infinite distance from the nucleon.  The remaining $2/3$ is
found at typical hadronic distances.  This surprising situation is
approached smoothly as $m_{\pi}\to 0$: A contribution to the axial
charge that approaches $g_{A}/3$ is found at distances of order
$1/m_{\pi}$ when $m_{\pi}$ is much greater than typical hadronic
distance scales.\cite{ref:1}

To demonstrate that this effect is physical, I compute the
effective $s$-wave spin-spin Hamiltonian for a hydrogenic atom
composed of a nucleon (mass $M$) and a charged lepton (mass
$m_{\ell}$) due to $Z^{0}$ exchange, for two cases, a) $M\gg
m_{\pi}\gg \alpha m_{\ell}$, and b) $M\gg \alpha m_{\ell}\gg
m_{\pi}$. As expected, in Case b) the interaction is exactly $2/3$ as
strong as
Case a).

Lattice estimates of the nucleon's axial vector charge extrapolated to
the chiral limit have always found values below the physical value of
$g_{A}\approx 1.26$.\cite{ref:2,history} These calculations are
performed relatively far from the chiral limit on relatively small
lattices.  Although my analysis is performed in the unbounded
continuum, it can be adapted to a finite domain and periodic boundary
conditions, and may lead to modification of lattice
extrapolations.\footnote{For an analysis of the implications of this
paper for lattice calculations see Ref.~\cite{Cohen:2001bg}.} Also,
estimates of $g_{A}$ in nuclei consistently fall below the value
obtained in isolation.\cite{ref:1a} The phenomenon discussed here may
account for medium dependent modifications of $g_{A}$ without invoking
partial restoration of chiral symmetry.

At first I work in the chiral limit, where the pion is massless and
$SU(2)_{L}\times SU(2)_{R}$ is an exact symmetry.  After deriving the
claimed result, I present a simpler, heuristic argument based on
symmetries, which may help develop the generalization to other
geometries.  Next, I consider the situation when the pion is light but
not massless, and show that the chiral limit is approached smoothly.
Finally I mention extensions to other hadronic axial charges and point
out that this anomaly is special to the axial current and does not
afflict either the nucleon's tensor charge or the matrix elements of
the tower of twist-two operators that determine its polarized quark
distribution, $\Delta q(x,Q^{2})$.

The nucleon's axial charge is defined by the matrix element of
the isovector axial current between momentum eigenstates:\cite{ref:3}
\begin{equation}
	\langle N(p',s')|\vec{\undertilde A}(0)|N( p,s)\rangle =\bar
	u(p',s')\fract{1}{2}\undertilde
	\tau\left(g_{A}(q^{2})\vec\gamma\gamma_{5}+h_{A}(q^{2})\vec
	q\gamma_{5}\right)u(p,s).
	\label{eq:1}
\end{equation}
The axial charge, $g_{A}$, is $g_{A}(0)$.  $h_{A}(q^{2})$ is the
induced pseudoscalar form factor, which receives a contribution from
the pion pole,
\begin{equation}
	h_{A}(q^{2})\equiv \frac{d_{A}(q^{2})}{q^{2}} =
	-\frac{2f_{\pi} g_{\pi N}}{q^{2}} +\cdots
	\label{eq:2}
\end{equation}
where $g_{\pi N}$ is the $\pi$-nucleon coupling constant, $f_{\pi}$ is
the pion decay constant, and the terms denoted by $\cdots$ are less
singular than $1/q^{2}$ as $q\to 0$.  Since the isovector axial
current is conserved in the chiral limit, $g_{A}$ and $g_{\pi N}$ are
related by the Goldberger-Treiman relation\cite{ref:4}
\begin{equation}
  	g_{A}=\frac{g_{\pi N}f_{\pi}}{M}
	\label{eq:3}
\end{equation}
which is exact when $m_{\pi}=0$ and approximate for small
$m_{\pi}$.\footnote{Other form factors,which are irrelevant for this
discussion, have been omitted from eq.~(\ref{eq:1}).}

Consider now the matrix element of the axial current in a nucleon
wavepacket state, $|N(s)\rangle$.  $N(s)$ represents a nucleon with
spin $\vec s$ at rest, centered at the origin.  It is described by a
wave packet normalized to unity, $|N(s)\rangle =\int
\frac{d^{3}p}{{(2\pi)}^{3/2}}\phi(p)|N(p,s)\rangle$, $\langle
N(s)|N(s)\rangle= \int d^{3}p\frac{E}{M}|\phi(p)|^{2}=1$.  Consider
the fourier transform of the current matrix element in the limit where
$|\vec q\ | $ is small compared to the momentum scale of the wave
packet.  Using eq.~(\ref{eq:1}) and the following properties of Dirac
spinors,
\begin{eqnarray}
	\bar u(p,s)\gamma_{\mu}u(p,s) = p_{\mu}/M \ ,\qquad \bar
	u(p,s)\gamma_{\mu}\gamma_{5}u(p,s) = s_{\mu}/M\nonumber\\
	    \bar u(p+q,s)\gamma_{5}u(p,s)\to \frac{1}{2M}
   q^{\mu}\bar u(p,s)\gamma_{\mu}\gamma_{5}u(p,s)\quad {\rm as } \
	q_{\mu}\to 0
	\label{eq:3a}
\end{eqnarray}
I obtain
\begin{equation}
	\lim_{\vec q\to 0}\int d^{3}r e^{i\vec q\cdot\vec r}
	\bigl\langle N(s)|\vec{\undertilde A}(\vec r)|N(s)\bigr\rangle
	= g_{A}\Bigl( \undertilde{\vec\chi} -\hat q\ \hat q\!\cdot \! 	
\undertilde {\vec\chi}\Bigr)
\label{eq:4}
\end{equation}
up to recoil corrections of order $\langle p^{2}\rangle/m^{2}$.
$\undertilde{\vec\chi}$ encodes the spin-isospin content of the matrix
element,
\begin{equation}
	\undertilde{\vec\chi} \equiv \fract{1}{2}
	U_{0}^{\dagger}\vec\sigma\undertilde{\tau}U_{0}.
	\label{eq:5}
\end{equation}
$U_{0}$ is a spinor-isospinor for a nucleon at rest.  In deriving
eq.~(\ref{eq:4}) I have used the Goldberger-Treiman relation to
express the entire matrix element in terms of $g_{A}$.  Conservation
of the axial current in the $\vec q\to 0$ limit is manifested by the
fact that $\hat q$ contracted with eq.~(\ref{eq:4}) vanishes.  In this
calculation it is necessary to treat the $\vec q\to 0$ limit
carefully.  Since $q^{0}\sim \vec q\ ^{2}/2M$ for small $\vec q$, I
have replaced $q^{2}$ by $-{\vec q\ }^{2}$.  Also from
eq.~(\ref{eq:3a}), $\bar u(q,s)\gamma_{5} u(0,s)\to \frac{1}{2M}
U_{0}^{\dagger} \vec\sigma\cdot\vec q\ U_{0}$ as $\vec q\to 0$.  The
term proportional to $\hat q\ \hat q\cdot\vec\sigma$ in
eq.~(\ref{eq:4}) shows that the limit $\vec q\to 0$ is nonuniform.

To separate the short- and long-range contributions to the matrix
element I define contributions to eq.~(\ref{eq:4}) from within and
outside a sphere of radius $R$ centered on the nucleon.  $R$ is taken
much greater than the range of typical QCD interactions, $R\gg
1/\Lambda$, and is held fixed as $\vec q\to 0$,
\begin{equation}
	\langle N|\vec{\undertilde A}(\vec r)|N\rangle =
	\theta(R-r)\undertilde{\vec a}{}_{0}(r)
	+\theta(r-R)\undertilde{\vec a}{}_{\pi}(r).
	\label{eq:7a}
\end{equation}
The first term is the unspecified contribution to the axial current
from typical hadronic distance scales.  For large enough $R$ only the
single pion contribution to the axial current remains when $r>R$, so
the second term is completely due to the pion, $\undertilde{\vec
A}\to-f_{\pi}\vec\nabla\undertilde\pi$, which in turn is given by the
familiar pole diagram.
\begin{equation}
	\undertilde{\vec a}{}_{\pi}(r)=\frac{g_{\pi N}f_{\pi}}{4\pi
	M}\vec\nabla\undertilde{\vec\chi}\cdot
	\vec\nabla\frac{1}{r}
	\label{eq:7b}
\end{equation}
Consider first the contribution of $\theta(R-r)\undertilde{\vec
a}{}_{0}$: Because the integral is bounded by $r\le R$, the $\vec q\to
0$ limit can be taken without any special precaution.  Therefore this
piece of the axial current can only contribute to $g_{A}$, not to the
pion pole.  Thus this contribution to eq.~(\ref{eq:4}) is given by
\begin{equation}
	\lim_{q\to 0}\int_{0}^{R}r^{2}dr d\Omega\ e^{i\vec q\cdot\vec
	r} {\undertilde {\vec a}}{}_{0}(\vec r) = \int_{0}^{R}r^{2}dr
	d\Omega\ {\undertilde {\vec a}}{}_{0}(\vec r)\equiv
	g_{A}^{0}(R)\undertilde{\vec\chi}.
	\label{eq:8}
\end{equation}
There is no need to specify $g_{A}^{0}(R)$ further, except to
emphasize that it is the result one would obtain by restricting the
calculation to the interior of a sphere of radius $R$.

Next turn to the region $r>R$, where the pion contribution to the
axial current is explicit and exact,
\begin{equation}
	\int_{R}^{\infty}r^{2}dr \oint d\Omega\ e^{i\vec q\cdot\vec r}
	{\undertilde {\vec a}}{}_{\pi}(\vec r)=
	-C\int_{R}^{\infty}r^{2}dr \oint d\Omega \ e^{i\vec q\cdot\vec
	r}\ \vec\nabla \Bigl(\frac{1}{r^{2}}\hat r\cdot
	\undertilde{\vec\chi}\Bigr)
	\label{eq:11}
\end{equation}
where $C=g_{\pi N}f_{\pi}/{4\pi M}$.  It remains merely to evaluate
this integral, mindful that $\vec q\to 0$ with $R$ fixed.

Note that dimensional analysis suggests that the contribution from
eq.~(\ref{eq:11}) will persist as $\vec q\to 0$ no matter how large
$R$ is, because the $r$ integration is scale invariant.  It must
generate the pion pole term, $-\hat q\ \hat
q\cdot\undertilde{\vec\chi}g_{A}$ in eq.~(\ref{eq:4}).  The surprise
is that there is another term proportional to $\undertilde{\vec\chi}$
alone, which contributes to $g_{A}$.

Eq.~(\ref{eq:11}) may be evaluated directly at fixed $R$ and $\vec
q$.  In particular, the oscillatory $\vec q$ dependence ensures
convergence at large $r$ and does not require any special treatment.
It is somewhat easier however to integrate by parts:
\begin{eqnarray}
	\int_{R}^{\infty}r^{2}dr \oint d\Omega\ e^{i\vec q\cdot\vec r}
	{\undertilde {\vec a}}{}_{\pi}(\vec r)&=& Ci\vec
	q\int_{R}^{\infty}dr \oint d\Omega e^{i\vec q\cdot\vec r}\
	\hat r\!\cdot \!\undertilde{\vec\chi} +C\left.\oint d\Omega
	e^{i\vec q\cdot \vec r}\hat r\ \hat r\!\cdot
	\!\undertilde{\vec\chi}\ \right|^{R}_{\infty}\nonumber\\
	&=&\undertilde{\vec\Gamma}{}_{V}+
	\undertilde{\vec\Gamma}{}_{ R}+\undertilde{\vec\Gamma}{}_{\infty},
\end{eqnarray}
where $\undertilde{\vec\Gamma}{}_{V}$ is the volume contribution for
$r\ge R$, and $\undertilde{\vec\Gamma}{}_{R, \infty}$ are surface
terms at $R$ and at the surface at infinity.

The surface at infinity does not contribute.  To see this, expand the
exponential using $\exp(i\vec q\cdot\vec r) = \sum_{\ell}i^{\ell}
(2\ell+1) P_{\ell}(\hat q\cdot\hat r)j_{\ell}(qr)$.  From the physical
definition of $g_{A}$ it is clear that $\undertilde
{\vec\Gamma}{}_{\infty}$ is to be evaluated on the surface at infinity
with $q$ and $R$ fixed.  Only $\ell\le 2$ contributes and the Bessel
functions $j_{\ell}(qr)$ go to zero as $r\to\infty$ at fixed $q$,
so $\undertilde{\vec\Gamma}{}_{\infty}=0$.

Next consider the volume contribution.  $\hat r\cdot
\undertilde{\vec\chi}$ selects out the $\ell=1$ term in the Bessel
expansion of the exponential:
\begin{eqnarray}
	\undertilde{\vec{\Gamma}}{}_{V} &=& -3 C\vec
	q\int_{R}^{\infty}dr j_{1}(qr)\int d\Omega\ \hat
	q\!\cdot\!\hat r\ \hat r\!\cdot\!  \undertilde{\vec\chi}
	\nonumber\\
	&=&-4\pi C\hat q\ \hat q\!\cdot\!\undertilde{\vec\chi}\
	j_{0}(qR)
	\label{eq:14}
\end{eqnarray}
where I have used $\langle \hat r_{i}\hat
r_{j}\rangle=\fract{1}{3}\delta_{ij}$.  Taking $\vec q\to 0$ at fixed
$R$, I obtain
\begin{eqnarray}
	\lim_{q\to 0}\undertilde{\vec{\Gamma}}{}_{V} =-\frac{g_{\pi
	N}f_{\pi}}{M}\ \hat q\ \hat q\!\cdot\!\undertilde{\vec\chi}
	\label{eq:15}
\end{eqnarray}
which is the expected pion pole term.

Finally consider the surface term at $R$.  With $R$ fixed and $\vec
q\to 0$, the exponential can be replaced by unity with the result
\begin{eqnarray}
	\lim_{q\to 0}\undertilde{\vec\Gamma}{}_{ R}&=&C\int d\Omega\ \hat
	r\ \hat r\!\cdot \!\undertilde{\vec\chi}\nonumber\\
	&=& \frac{g_{\pi N}f_{\pi}}{3M}\undertilde{\vec\chi}
	\label{eq:13}
\end{eqnarray}
which contributes to $g_{A}$.  Using the Goldberger-Treiman relation
the magnitude of this term is $g_{A}/3$ independent of $R$.

The three contributions to eq.~(\ref{eq:4}) given in
eqs.~(\ref{eq:8}), (\ref{eq:15}), and (\ref{eq:13}) combine to give
\begin{equation}
	\lim_{\vec q\to 0}\int d^{3}r e^{i\vec q\cdot\vec r} \langle
	N(s)|\vec{\undertilde A}(\vec r)|N(s)\rangle=
	g_{A}^{0}(R)\undertilde{\vec\chi} +\frac{1}{3}
	g_{A}\undertilde{\vec\chi} - g_{A} \ \hat q\ \hat
	q\!\cdot\!\undertilde{\vec\chi}
	\label{eq:16}
\end{equation}
again using the Goldberger-Treiman relation.

Comparison with eq.~(\ref{eq:4}) verifies that we have obtained the
correct pion pole contribution (a check on the algebra) and that the
contribution to $g_{A}$ from inside an arbitrary but fixed radius $R$
equals $\frac{2}{3}g_{A}$ independent of $R$.

Since this result is unexpected it is worth presenting a somewhat
different argument, which is simpler, but heuristic.  This argument
may generalize to other geometries.  If one suspects a possible long
range contribution to the matrix element on the left-hand side of
eq.~(\ref{eq:4}) one can suppress it by averaging over the angles of
$\vec q$ before sending $\vec q\to 0$.  This has the effect of
replacing $\exp(i\vec q\cdot\vec r)$ by $\sin qr/qr$, which is
suppressed by an additional power of $r$ at large $r$.  Thus the
spherical average of eq.~(\ref{eq:4}) is equivalent to the short-range
contribution piece of eq.~(\ref{eq:4}) and should be identified with
$g_{A}^{0}(R) \undertilde{\vec\chi}$.  A factor of $1/3$ appears when
one averages $\hat q\ \hat q\!\cdot\!\vec \sigma$ over the sphere,
with the result
\begin{equation}
	\lim_{\vec q\to 0}\biggl\langle\int d^{3}r e^{i\vec q\cdot\vec
	r} \langle N(s)|\vec{\undertilde A}(\vec r)|N(s)\rangle
	\biggr\rangle=g_{A}^{0}(R) \undertilde{\vec\chi}= \fract{2}{3}
	g_{A} \undertilde{\vec\chi}
	\label{eq:17}
\end{equation}
which reproduces our earlier result.

Thus I conclude that in the chiral limit the correct value of the
nucleon's axial charge is $3/2$ the value that is computed by
restricting the calculation to any finite volume.

Next, I repeat the analysis with a nonzero pion mass and study the
behavior of $g_{A}$ as a function of $m_{\pi}$ and $R$.  Although the
form of the analysis changes considerably, the physical result changes
smoothly as one departs from the chiral limit.  There is no longer a
pion pole in the induced pseudoscalar form factor, so that term plays
no role in the analysis as $\vec q\to 0$.  Note also that the
Goldberger-Treiman relation is no longer exact so I will not replace
$g_{\pi N}f_{\pi}/M$ by $g_{A}$ except in the limit $m_{\pi}\to 0$.
All contributions to the matrix element of $\undertilde{\vec A}(r)$
fall exponentially, $\sim e^{-m_{\pi}r}$, for large $r$.  Still, for
small $m_{\pi}$ and large $r$ ($m_{\pi}, 1/r\ll \Lambda$) the pion
dominates and its contribution can be calculated analytically.  In
place of eq.~(\ref{eq:7b}) one has
\begin{equation}
	\undertilde{\vec a}{}_{\pi}(\vec r)
	=\frac{g_{\pi N}f_{\pi}}{4\pi M} \vec\nabla\undertilde{\vec\chi}
	\cdot\vec\nabla\Bigl(\frac{e^{-m_{\pi}r}}{r}\Bigr).
	\label{eq:18}
\end{equation}
It is no longer necessary to be so careful about the limit $\vec q\to
0$.  Instead the matrix element of the axial current at $\vec q=0$ can
be calculated directly.  It can be separated into the contribution
from $r\le R$, again labelled $g_{A}^{0}(R)$, and the contribution
from $r>R$ where the pion dominates.  A straightforward calculation
yields
\begin{equation}
	\int d^{3}r \langle N(s)|\vec{\undertilde A}(\vec
	r)|N(s)\rangle= g_{A}\undertilde{\vec\chi}=
	\Bigl(g_{A}^{0}(R)+\frac{1}{3}\frac{g_{\pi N}f_{\pi}}{M}
	(1+m_{\pi}R)e^{-m_{\pi}R} \Bigr)\undertilde{\vec\chi}.
	\label{eq:19}
\end{equation}
This result displays the expected behavior: For any fixed $R$, as
$m_{\pi}\to 0$, one third of the axial charge comes from outside $R$.
However, for any fixed $m_{\pi}$, as $R\to\infty$ all the axial charge
comes from inside $R$.  Finally, it is possible to calculate the
density of axial charge in the region where the pion dominates.  The
result,
\begin{equation}
	\frac{dg_{A}}{dr}=\frac{1}{3}\frac{g_{\pi
	N}f_{\pi}}{M}m_{\pi}^{2}re^{-m_{\pi}r}
	\label{eq:20}
\end{equation}
valid at large $r$, again displays the nonuniformity of the
$m_{\pi}\to 0$ and $r\to\infty$ limits.  Whether these results are
modified by chiral logarithms for small but nonvanishing $m_{\pi}$ is
outside the scope of this Letter.

A comment is in order on the relation of this result to traditional
calculations of the pionic contribution to the nucleon's axial
charge.\cite{Bernard:2001rs}  It is well known that, when integrated
{\it over all space\/}, the pion contributes to $h_{A}(q^{2})$ and
not to $g_{A}(q^{2})$.  This can be seen by Fourier transforming
eq.~(\ref{eq:18}) with the result,
\begin{equation}
   \int d^{3}r e^{i\vec q\cdot\vec r}
   \bigl\langle N(s)|\undertilde{\vec a}{}_{\pi}(\vec
r)|N(s)\bigr\rangle
   = -\frac{g_{\pi N}f_{\pi}/M}{\vec q^{2}+m_{\pi}^{2}}
    \vec q\undertilde{\vec\chi}\cdot\vec q\  ,
   \label{eq:21}
\end{equation}
(assuming some smearing of the pion source at the scale of the nucleon
size).  The tensor structure $\vec q\undertilde{\vec\chi}\cdot\vec q$
signals a contribution to $h_{A}(-\vec q{\ }^{2})$ and no contribution
to $g_{A}(-\vec q{\ }^{2})$.  There is no puzzle here: we are studying
the contribution to the axial charge outside a sphere of radius $R$
surrounding a localized nucleon, not the integral over all space.  The
result we have derived is independent of the nature of the short range
contributions to the nucleon's axial charge and the fact that the
total pion pole contribution vanishes.  Finally, the reader may wonder
what additional short range term appears in the pion contribution to
$dg_{A}/dr$ (see eq.~(\ref{eq:20})) to make the integral over $r$
vanish.  A study of eq.~(\ref{eq:18}) at small $r$ reveals the
presence of a Dirac $\delta$ function, so the {\it pionic\/}
contribution to $dg_{A}/dr$ is %
\begin{equation}
   \left.\frac{dg_{A}}{dr}\right|_{\rm
   pion}=\frac{1}{3}\frac{g_{\pi N}f_{\pi}}{M}\left(m_{\pi}^{2}r
   e^{-m_{\pi}r}-2\delta(r)\right)\ ,
\end{equation}
which integrates to zero.  For an extended
nucleon source, presumably the $\delta$ function would be replaced by
a smooth contribution with the same integral.

One might worry that the delocalization of the axial charge
demonstrated here is a merely a mathematical oddity rather than a
physically observable effect.  To put this doubt to rest, consider the
following {\it gedanken\/} experiment in which it could, in principle,
be measured.  Consider a hydrogenic atom composed of a proton bound to
a lepton of mass $m_{\ell}$.  We take $m_{\ell}$ small enough that the
Bohr radius, $a_{\ell}=1/m_{\ell}\alpha$, is much greater than
$1/\Lambda_{\rm QCD}$, the range of the strong interactions, excluding
the pion which we treat explicitly.  Consider the proton and lepton
interaction mediated by $Z^{0}$ exchange.  The axial-axial piece of
$Z^{0}$ exchange generates a spin-spin interaction which contributes a
very small (but in principle observable) correction to the hydrogen
hyperfine splitting.  An elementary calculation yields,
\begin{equation}
   H_{Z}(\vec r)=-\frac{G_{F}}{\sqrt{2}}\langle P(s)|
   \vec A_{3}(\vec r)|P(s)\rangle\cdot\vec\sigma_{\ell}
   \label{Zint}
\end{equation}
where $|P(s)\rangle$ is a proton with spin $\vec s\,$ localized at
the origin, $\vec A_{3}$ is the third component of $\undertilde{\vec
A}$ in isospin space, and $\half\vec\sigma_{\ell}$ is the lepton spin
operator.
We are interested in the chiral limit, so we take $m_{\pi}\ll
\Lambda_{\rm QCD}< M$.  We parameterize $\undertilde{\vec A}$ in
terms of $\undertilde{\vec a}{}_{0}$ and $\undertilde{\vec
a}{}_{\pi}$ (see
eqs.~(\ref{eq:7a}) and (\ref{eq:18}).  In the $s$-wave only the
spherical average of ${\vec A}_{3}$ survives, and we find,
\begin{equation}
   \langle\langle\vec A_{3}(r)\rangle\rangle
   =\left(\frac{2}{3}g_{A}\delta^{3}(\vec r)
   +\frac{1}{12\pi}g_{A}m_{\pi}^{2}\frac{
   e^{-m_{\pi} r}}{r}\right)\half\vec\sigma_{P},
   \label{sphericalave}
\end{equation}
where $\langle\langle\vec A_{3}(r)\rangle\rangle$ denotes the
spherical average of the proton matrix element of $\vec A_{3}(\vec
r)$.  I have replaced the short range contribution, $\undertilde {\vec
a}{}_{0}(r)\theta(R-r)$ by $g_{A}^{0}\undertilde{\vec\chi}
\delta^{3}(\vec r)$ since the hydrogenic ground state wave function is
essentially constant for $r\le R$.  Also, in averaging the pionic
contribution I have omitted the delta function at the origin in
$\nabla^{2}(e^{-m_{\pi}r}/r)$ because the origin is explicitly
excluded by the $\theta(r-R)$.  Combining eqs.~(\ref{Zint}) and
(\ref{sphericalave}), yields an effective Hamiltonian for the
spin-spin interaction in the $s$-wave,
\begin{equation}
   H_{s{\rm -wave}}(r)=-\frac{G_{F}}{2\sqrt{2}}
   \left(\frac{2}{3}g_{A}\delta^{3}(\vec r)
   +\frac{1}{12\pi}g_{A}m_{\pi}^{2}\frac{
   e^{-m_{\pi} r}}{r}\right)\vec\sigma_{P}
   \cdot\vec\sigma_{\ell}
\end{equation}
Next, take the expectation value of $H_{s{\rm -wave}}(\vec r)$ in the
ground state $\psi_{0}(r)=(\fract{1}{\pi
a_{\ell}})^{3/2}e^{-r/a_{\ell}}$.  If $1/m_{\pi}\gg a_{\ell}$, then
the second term in eq.~(\ref{sphericalave}) is negligible and we
obtain,
\begin{equation}
   \lim_{m_{\pi}a_{\ell}\to 0}\langle\psi_{0}|H_{s{\rm -wave}}
   |\psi_{0}\rangle =-\frac{G_{F}}{2\sqrt{2}}
   \left(\frac{2}{3}g_{A}\right)\vec\sigma_{P}\cdot \vec\sigma_{\ell}
   |\psi_{0}(0)|^{2}
   \label{chiral}
\end{equation}
which verifies that one-third of the nucleon's axial charge has
receded to distances of order $1/m_{\pi}$, which is much greater than
the size of the atom and therefore does not affect the interaction.
On the other hand, if $1/m_{\pi}\ll a_{\ell}$ then the second term in
eq.~(\ref{sphericalave}) is easily evaluated and supplies the missing
third of $g_{A}$:
\begin{equation}
   \lim_{m_{\pi}a_{\ell}\to \infty}\langle\psi_{0}|H_{s{\rm -wave}}
   |\psi_{0}\rangle =-\frac{G_{F}}{2\sqrt{2}}
   g_{A}\vec\sigma_{P}\cdot \vec\sigma_{\ell}
   |\psi_{0}(0)|^{2}
   \label{unchiral}
\end{equation}
The latter case goes smoothly over to the physically relevant case
(with strength $g_{A}$) in which $m_{\pi}$ and $\Lambda_{\rm QCD}$ are
comparable and much greater than $\alpha m_{\ell}$.\footnote{Note that
in this calculation it is justified to ignore the $\vec q\,{}^{2}$
variation of the axial form factor because the typical momentum
transfer in the hydrogenic ground state is determined by $m_{\ell}$,
which can be taken to be small compared to the natural scale of
variation of the axial form factor (which is finite in the chiral
limit).\cite{Bernard:2001rs} Note also that I have identified $g_{\pi
N}f_{\pi}/M$ with $g_{A}$ even for $m_{\pi}\ne 0$, which is only
approximately true.} This example shows that the delocalization of the
axial charge in the chiral limit is (at least in principle) a
physically measurable effect.

Other hadron axial charges, for example, the ones that enter
strangeness conserving semileptonic hyperon decays (like
$\Sigma\to\Lambda e \bar \nu$) or semileptonic $D$ and $B$ decays
(like $B^{*}\to B e\bar\nu$), behave in the same way as $m_{\pi}\to
0$.  However, it appears that other nucleon matrix elements of
pseudotensor operators {\it do not\/} suffer from this subtlety that
afflicts the axial charge.  The isovector tensor charge, $g_{T}$,
defined as the coefficient of $p_{\mu}s_{\nu}-p_{\nu}s_{\mu}$ in the
forward matrix element of $\bar \psi \sigma_{\mu\nu}\gamma_{5}
\fract{1}{2} \undertilde\tau \psi$, provides an important contrast
because it is amenable to lattice simulation.  There is no pion pole
term in the tensor charge because $\langle \pi(p)|\bar \psi
\sigma_{\mu\nu}\gamma_{5}\fract{1}{2} \undertilde\tau \psi|0\rangle
=0$.  So lattice calculations of $g_{T}$ should be more stable in the
chiral limit than $g_{A}$.  Another important tower of operators are
the twist-two, pseudotensor operators that determine the
spin-dependent quark distributions in the nucleon, ${\cal
A}^{n}_{\mu_{1}\mu_{2}\cdots\mu_{n}}=\bar \psi\gamma_{\{\mu_{1}}
D_{\mu_{2}}\cdots D_{\mu_{n}\}}\gamma_{5}\psi$.  ${\cal A}^{n}$ does
receive a pion pole contribution, but it is proportional to a high
(second or greater) derivative of $\undertilde \pi$, and therefore
does not generate a nonuniformity in the matrix element.

It appears that the flavor nonsinglet axial charge is unique in
possessing such singular behavior in the chiral limit.  This may be
the source of instabilities in lattice simulations of $g_{A}$.  The
unusual spatial distribution of the nucleon's axial charge may have
other physical consequences.  For example, that it receives an
important contribution from distance scales of order $1/m_{\pi}$ may
account for modifications of $g_{A}$ in large nuclei without invoking
partial restoration of chiral symmetry.

\section*{Acknowledgments}
I would like to thank T.~Blum, T.~Cohen, R.~Jackiw, J.~Negele,
S.~Ohta,
K.~Orginos, K.~Rajagopal, F.~Wilczek, and M.~Wise for conversations on
this subject.  This work is supported in part by the U.S.~Department
of Energy (D.O.E.) under cooperative research agreement
\#DF-FC02-94ER40818, and in part by the RIKEN-BNL Research Center at
Brookhaven National Laboratory.


\end{document}